\documentclass[aps,prb,twocolumn,superscriptaddress]{revtex4-2}
\usepackage{graphicx}
\usepackage{mathrsfs}
\usepackage{bm}
\usepackage{amsmath}
\usepackage{color}
\usepackage{dcolumn}
\usepackage{epstopdf}
\usepackage{dsfont}
\usepackage{amssymb}
\usepackage{tabularx}
\usepackage{array,ulem}
\usepackage{float}
\usepackage{colordvi}
\usepackage{verbatim}
\usepackage{hyperref}
\hypersetup{
	colorlinks=true,
	linkcolor=blue,
	filecolor=blue,
	urlcolor=blue,
	citecolor=blue,
}

\begin{document}
\title{Two-dimensional higher-order Weyl semimetals}

\author{Lizhou Liu}
\affiliation{College of Physics, Hebei Normal University, Shijiazhuang 050024, China}

\author{Qing-Feng Sun}
\email[Correspondence author:~~]{sunqf@pku.edu.cn}
\affiliation{International Center for Quantum Materials, School of Physics, Peking University, Beijing 100871, China}
\affiliation{Hefei National Laboratory, Hefei 230088, China}

\author{Ying-Tao Zhang}
\email[Correspondence author:~~]{zhangyt@mail.hebtu.edu.cn}
\affiliation{College of Physics, Hebei Normal University, Shijiazhuang 050024, China}

\date{\today}

\begin{abstract}
We propose a theoretical scheme to realize two-dimensional higher-order Weyl semimetals using a trilayer topological insulator film coupled with a $d$-wave altermagnet. Our results show that the trilayer topological insulator exhibits two-dimensional Weyl semimetal characteristics with helical edge states. Notably, the Weyl points are located at four high-symmetry points in the Brillouin zone, and the topology of symmetric subspaces governs the formation of these Weyl points and edge states. Upon introducing a $d$-wave altermagnet oriented along the $z$-direction, gaps open in the helical edge states while preserving two Weyl points, leading to the realization of two-dimensional higher-order Weyl semimetals hosting topological corner states. The nonzero winding number in the subspace along the high-symmetry line serves as a topological invariant characterizing these corner states, and the other subspace Hamiltonian confirms the existence of the Weyl points.
Finally, a topological phase diagram provides a complete topological description of the system.
\end{abstract}

\maketitle

\maketitle
\section{Introduction}

 In recent years, the study of topological phases has expanded significantly, uncovering novel properties across a wide range of systems~\cite{Chiu2016, Bansil2016}.
Beyond the traditional bulk-boundary correspondence observed in topological insulators~\cite{Hasan2010, LiuFeng, Qi2011, Haldane1988, Kane2005, Kane2005a, Bernevig2006, Bernevig2006a} and semimetals~\cite{Armitage2018, Lv2021, Burkov2016, Dai2016}, the emergence of higher-order topological phases has marked a paradigm shift.
Initially proposed for insulators, these higher-order topological phases exhibit distinctive bulk-boundary relations, including corner states in two- or three-dimensional systems and hinge states in three-dimensional systems~\cite{Benalcazar2017, Benalcazar2017a, Li2020, Benalcazar2019, Schindler2018, Peterson2018, Yao2023, Bhowmik2024, Liu2023, Hung2024, Mazanov2024}.
A major research focus has been the induction of higher-order corner states by breaking the edge states of two-dimensional (2D) topological insulators~\cite{Ren2020, Zhuang2022, Han2022, Miao2022, Miao2023, Miao2024, Chen2020a, Li2024, Liu2024}.
Building on these 2D corner states, second-order topological semimetal states with hinge states, such as Weyl semimetals, Dirac semimetals, and nodal ring semimetals, have been predicted in bulk-closed three-dimensional systems~\cite{Lin2018, Ghorashi2020, Wang2020b, Xiong2023, Pu2023, Ghorashi2021, Wang2022, Chen2022, Gao2023, Du2022, Hirsbrunner2024, Pan2024, Qi2024}.
{The dimension reduction may produce unique physical properties, such as parity anomaly in $(2 + 1)$-dimensional (space-time) quantum field theory~\cite{Jackiw1984, Fradkin1986, Semenoff1984, Mogi2022}, giant Berry curvature dipole~\cite{Sodemann2015, Du2021}, and topological quantum criticality~\cite{Ezawa2012}.
Moreover, compared to hinge states, corner states in higher-order systems are highly manipulable~\cite{Han2024}, and their association with semimetals may give rise to more exotic physical phenomena.}
However, the connection between higher-order topology, including corner states, and 2D semimetals remains an open question.

Recently, altermagnetism, distinct from both ferromagnetism and antiferromagnetism, has been proposed within the framework of spin group theory~\cite{Smejkal2022, Smejkal2022a, Smejkal2022b}.
 In altermagnetic systems, sublattices with opposite spins are not related by spatial inversion or fractional translation operations but are instead connected by rotation or mirror symmetry operations.
 This unique symmetry constraint gives rise to non-relativistic, anisotropic spin splitting in the Brillouin zone.
As a result, altermagnets-combining real-space antiferromagnetic order with momentum-space anisotropic spin splitting, present a promising platform for realizing 2D higher-order semimetals.

\begin{figure}
  \centering
  \includegraphics[width=8.5cm,angle=0]{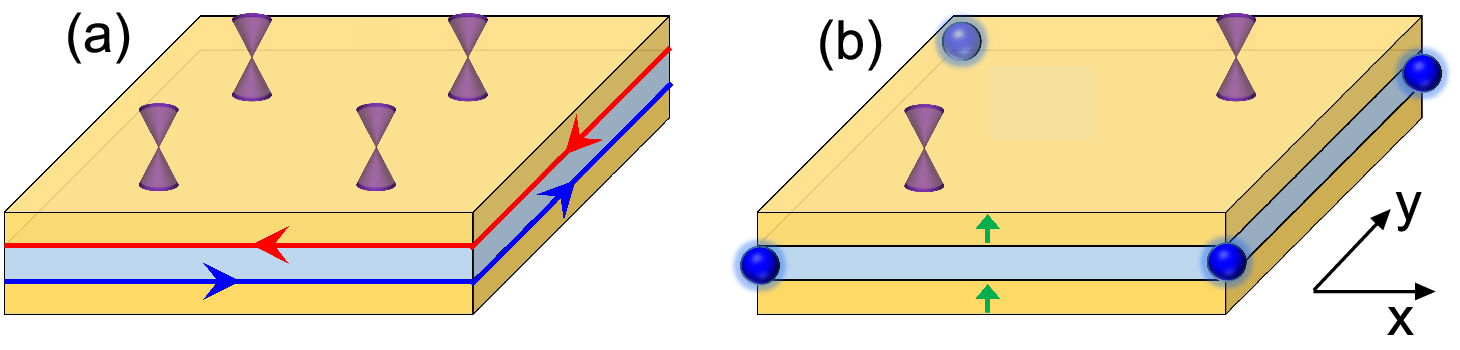}
\caption{(a) Schematic of a trilayer topological insulator film. The films are represented in yellow and light blue layer, respectively.
Red and blue lines indicate helical edge states.
(b) Schematic of a trilayer film with $d$-wave altermagnet. The green arrows indicate the direction of the altermagnet parallel to the out-of-plane.
The purple cones indicate the Weyl cones in the first Brillouin zone and the blue spheres indicate the corner states.
}
  \label{fig1}
\end{figure}

In this work, we theoretically propose a pioneering scheme to realize 2D higher-order Weyl semimetals with corner states, utilizing trilayer topological insulator films and $d$-wave altermagnets.
As illustrated in Fig.~\ref{fig1}(a), the 2D system hosts helical edge states and four bulk Weyl points at high symmetry points.
Introducing a $d$-wave altermagnet along the $z$-direction opens a gap in the helical edge states while preserving two bulk Weyl points.
 The emergence of corner states within the edge gaps signifies the realization of 2D higher-order Weyl semimetals, as shown in Fig.~\ref{fig1}(b).
 The topological invariants of the symmetric subspace characterise the topological nature of the system.
Finally, a topological phase diagram is determined.

\section{System Hamiltonian}
We adopt the system of Bi$_2$Se$_3$ topological insulator films~\cite{Wang2014, Lu2010, Li2020a}, the Hamiltonian for the trilayer films as shown in Fig.~\ref{fig1} can be expressed as follows:
\begin{align}
 H(k)&=\upsilon_F (\sin k_y s_x-\sin k_x s_y) \sigma_z + m (k)s_0 \sigma_x,
   \label{eq1}
\end{align}
where the first term corresponds to the kinetic energy, $\upsilon_F$ is the Fermi velocity, and the second term is the coupling term between the different layers, $m(k)=m_0 + 4 m_1 - 2 m_1 \cos k_x   - 2 m_1 \cos k_y $,  $m_0$ and $m_1$ are the hybridization gap and parabolic band components, respectively. Here $\sigma_{x, z}$ denote the layer degrees of freedom, and the three degrees of freedom correspond to the top, middle, and bottom layers, the concrete form is
\begin{align}
\sigma_z&=
\begin{bmatrix}
1 & 0 & 0 \\
0 & -1 & 0 \\
0 & 0 & 1
\end{bmatrix}, \qquad
\sigma_x=\begin{bmatrix}
0 & 1 & 0 \\
1 & 0 & 1 \\
0 & 1 & 0
\end{bmatrix}.
\end{align}
And $s_{x, y, z}$ is the Pauli matrices for the spin. The system is assumed to be half-filled with the Fermi level at zero energy.

The spin splitting of altermagnets exhibits various forms, including $p$-wave, $d$-wave, $g$-wave, and $i$-wave~\cite{Smejkal2022b}.
We employ $d$-wave spin splitting altermagnets, as follows
\begin{align}
  H_{AM}= J (\cos k_x  - \cos k_y) s_z  \otimes \begin{bmatrix}
1 & 0 & 0 \\
0 & 0 & 0 \\
0 & 0 & 1
\end{bmatrix}.
 \label{eq2}
\end{align}
Here $s_z$ represents altermagnet in the $z$-direction, and only the top and bottom layers are affected by altermagnet, which can be induced by magnetic proximity effect~\cite{Vobornik2011}.

\begin{figure}
  \centering
  \includegraphics[width=8.5cm,angle=0]{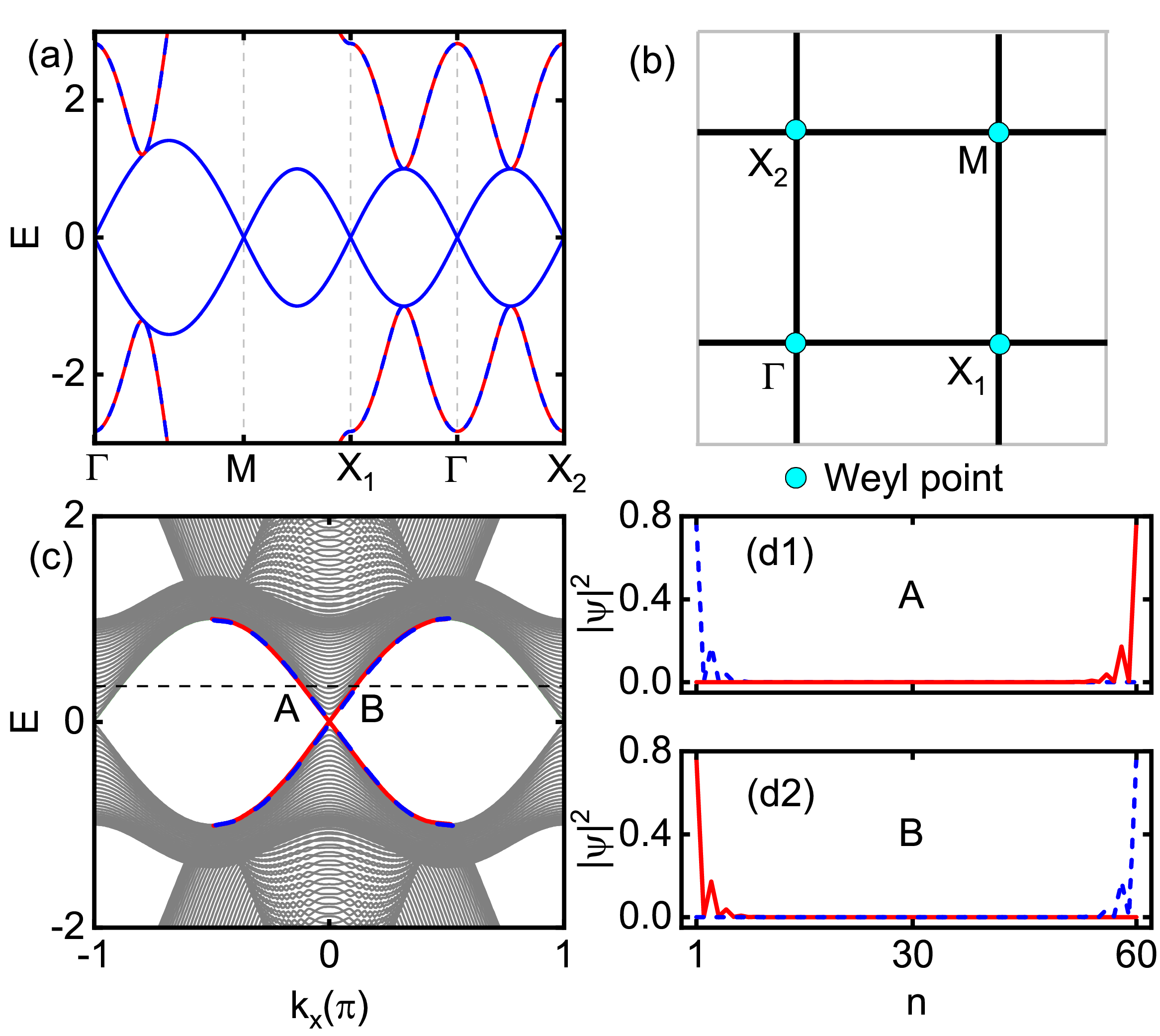}
 \caption{(a) Bulk energy band structure of the trilayer film along the high symmetry points. The bands far from the Fermi energy are twofold degenerate, and the bands near the Fermi energy are not degenerate.
 (b) Locations of the four Weyl points in the first Brillouin zone, with Weyl points indicated by cyan.
 (c) Energy band structure of nanoribbons. The solid red and dashed blue lines indicate gapless helical edge states.
  (d1) and (d2) denote the probability distributions of the edge states for $E= 0.2$ in A and B of (c), respectively.
  The blue and red edge states on the same boundary propagate in opposite directions, hence the overall helical edge state.
The parameters are chosen as $v_F = 1$, $m_1= 1$, $m_0 = -2$, and $J=0$.}
  \label{fig2}
\end{figure}

\section{Results and discussion}

\subsection{Weyl semimetal with helical edge state}

Let us first examine the system without the altermagnet $(J=0)$. The bulk band structure of the trilayer films along the high-symmetry points is shown in Fig.~\ref{fig2}(a).
One can see that the bulk bands far from the Fermi energy are twofold degenerate, while those bulk bands near the Fermi energy are non-degenerate and close at the high-symmetry points.
It indicates that the Weyl points in the 2D system are fixed at these high-symmetry points, and are not cleaved from Dirac point because of the lack of a spatial inversion counterpart.
The positions of the Weyl points (highlighted in cyan) within the Brillouin zone are shown in Fig.~\ref{fig2}(b).

To further investigate the edge states, we calculate the energy spectrum of nanoribbons as a function of $k_x$, with the $y$-direction subjected to open boundary conditions $(N_y = 50a)$, where $a$ denotes the lattice constant.
Interestingly the 2D Weyl semimetal exhibits edge states, as illustrated by the red solid and blue dashed lines in Fig.~\ref{fig2}(c).
To characterize these states more precisely, we analyze the wave function distribution corresponding to the edge states at energy $E = 0.2$, as shown in Fig.~\ref{fig2}(d1) and Fig.~\ref{fig2}(d2).
We observe that the two edge states at point $A$ are localized at opposite ends of the nanoribbon.
Similarly, the edge states at point $B$ are also distributed at opposite ends, but their spatial distribution is reversed compared to those at point $A$.
This spatial distribution confirms that these are helical edge states.
Thus, the above analysis establishes that the trilayer film system is a 2D Weyl semimetal with helical edge states.

To elucidate the physical origin of the Weyl points and helical edge states, we examine the symmetry of the bulk Hamiltonian $H(k)$.
In the absence of altermagnet, the Hamiltonian $H(k)$ retains invariance under the symmetry $\mathcal{M}$, where
\begin{align}
  \mathcal{M}= \frac{\sqrt{2}}{2}\begin{bmatrix}
0 & s_z & 0 \\
s_z & 0 & s_z \\
0 & s_z & 0
\end{bmatrix}.
 \label{eq3}
\end{align}
It is straightforward to find that $\mathcal{M}$ has three eigenvalues $\pm 1$ and $0$.
The eigenvectors of subspace with eigenvalue $-1$ are $[0, -\frac{1}{2},0,  -\frac{\sqrt{2}}{2}, 0 ,- \frac{1}{2}]^T$ and $[\frac{1}{2}, 0,-\frac{\sqrt{2}}{2}, 0 , \frac{1}{2},0]^T$,
the eigenvectors of subspace with eigenvalue $+1$ are $[\frac{1}{2}, 0,\frac{\sqrt{2}}{2}, 0 , \frac{1}{2},0]^T$ and $[0, -\frac{1}{2},0,  \frac{\sqrt{2}}{2}, 0 ,- \frac{1}{2}]^T$,
and the eigenvectors of subspace with eigenvalue $0$ are $[\frac{\sqrt{2}}{2}, 0,0, 0 , -\frac{\sqrt{2}}{2},0]^T$ and $[0, -\frac{\sqrt{2}}{2},0,  0, 0 , \frac{\sqrt{2}}{2}]^T$.
In these three subspaces, $H(k)$ can be decoupled into three parts:
\begin{align}
H_{\pm1}&=- v_F \sin k_y \tau_x \pm v_F \sin k_x \tau_y + \sqrt{2} m(k) \tau_z,\nonumber \\
H_{0}&=- v_F \sin k_y \tau_x +v_F \sin k_x \tau_y \equiv \textbf{d}_{0}(k) \cdot {\bf{\tau}} ,  \label{eq4}
\end{align}
where ${\bf {\tau}}=(\tau_{x},\tau_y,\tau_z)$ are the Pauli matrix. Each of $H_{-1}, H_{1},$ and $H_{0}$ exhibits chiral symmetry.
Notably, $H_0$ is zero at the four high symmetry points $\Gamma (0,0), X_1 (\pi,0), X_2 (\pi,0),$ and $M (\pi,\pi)$, which correspond to the four Weyl points, as shown in Fig.~\ref{fig2}(b).
We further investigate the topological charge associated with each Weyl point in the subspace Hamiltonian $H_0$.
The winding number for each Weyl point is calculated using the following expression~\cite{Chiu2016}:
\begin{align}
  v_{W}=\frac{1}{2 \pi} \int_C (\textbf{d}_0(k) \times \frac{d}{dk} \textbf{d}_0(k) )d k,
 \label{eq6}
\end{align}
where $W=\Gamma, X_1, X_2, M$ and $C$ denotes an enclosing loop around a Weyl point in the 2D Brillouin zone.
It follows that $v_\Gamma = v_M = - v_{X_1}=-v_{X_2}=1$, corresponding to the four Weyl points.
The total topological charge of the two Weyl points along the $k_x = 0$ and $k_x = \pi$ projections is zero, $i.e.,$ $v_\Gamma + v_{X_2} = 0$ and $v_M + v_{X_1} = 0$.
As a result, the energy band structure of the nanoribbons, shown in Fig.~\ref{fig2}(c), does not exhibit one-dimensional Fermi arcs connecting the Weyl points.

The Chern numbers of bulk gapped $H_1$ and $H_{-1}$ can be calculated using the following expression~\cite{Thouless1982, Xiao2010, Matthes2016}
\begin{eqnarray}
\mathcal{C}=\frac{1}{2 \pi}\sum_n \int_{BZ} d^2 \mathbf{k} \Omega_n,
\label{EQ4}
\end{eqnarray}
where $\mathbf{k} =(k_x,k_y)$ and $\Omega_n$ is the momentum-space Berry curvature for the $n$th band~\cite{Qiao2018, Yao2004, Chang1996}
\begin{eqnarray}
\Omega_n(\mathbf{k})= -\sum_{m \neq n}  \frac{2{\rm  Im} \langle \psi_{n \mathbf{k}} | v_x | \psi_{m \mathbf{k}} \rangle \langle  \psi_{m \mathbf{k}} | v_y | \psi_{n \mathbf{k}} \rangle }{ (\omega_m - \omega_n)^2}.
\label{EQ5}
\end{eqnarray}
In this expression, the sum runs over all occupied bands below the bulk gap, where $\omega_n = E_n / \hbar$ represents the energy of the $n$-th band, and $v_{x,y}$ are the velocity operators.
By integrating the Berry curvature, we calculate the Chern numbers and find $\mathcal{C}_{H_1} = - \mathcal{C}_{H_{-1}} = 1$, which corresponds to a pair of the opposite chiral edge states in open boundary, i.e., helical edge states.
These results indicate that the Hamiltonians of the three subspaces collectively describe a 2D Weyl semimetal with helical edge states.

\subsection{Corner states in Weyl semimetal}

\begin{figure}
  \centering
  \includegraphics[width=8.5cm,angle=0]{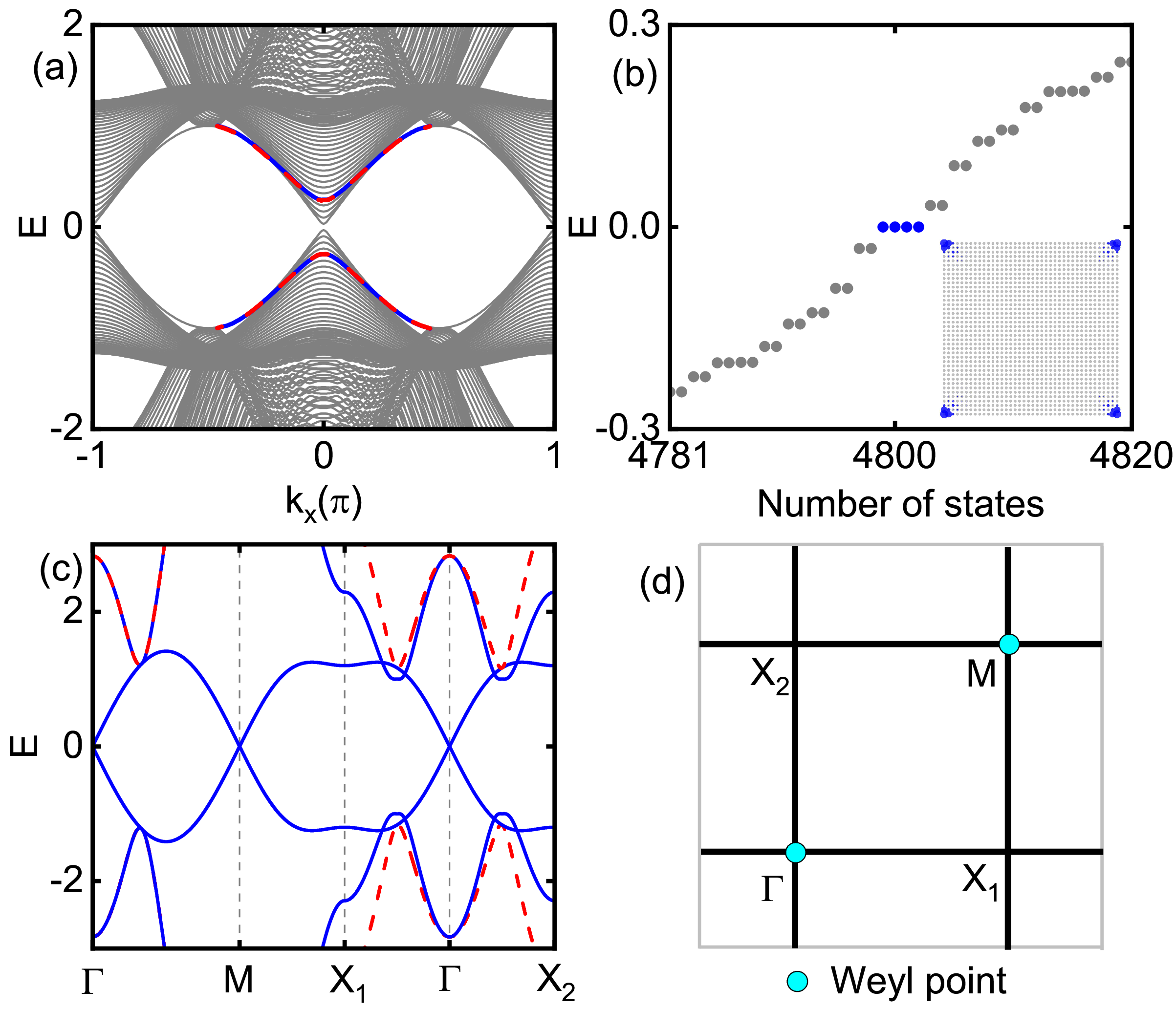}
\caption{(a) Band structures of nanoribbons of the topological insulator film with $d$-wave altermagnet. The blue solid line and the red dashed lines indicate the gapped helical edge states.
(b) Energy levels of the rectangular-shaped nanoflake. Blue dots correspond to corner states. The wave function probability distribution of the corner states is shown in the inset (blue density dots).
 (c) Bulk energy spectrum along the high symmetry points.
 (d) Locations of the two Weyl points in the first Brillouin zone, with Wely points indicated by cyan.
 The altermagnet is set to $J = 0.6$ and the other parameters are the same as in Fig.~\ref{fig2}.}
\label{fig3}
\end{figure}

We now introduce a $d$-wave altermagnet $(J = 0.6)$ to demonstrate the emergence of corner states.
The energy spectrum of the nanoribbon is shown in Fig.~\ref{fig3}(a), with the helical edge states being gapped by the altermagnetism (indicated by the solid blue and dashed red lines).
Interestingly, the bulk bands remain unaffected.
Despite the disappearance of the first-order edge states, we observe the emergence of second-order topological phases.
To investigate corner states, we analyze the energy spectrum of finite-size nanoflakes, such as a rectangular nanoflake with dimensions $40a \times 40a$.
As shown in Fig.~\ref{fig3}(b), four zero-energy states appear at the Fermi level (highlighted by blue dots) within the continuum spectrum.
The inset shows the wave-function probabilities of these corner states at half-filling, revealing that the electronic charge is localized at the corners of the nanoflake.

The energy bands shown in Fig.~\ref{fig3}(a) confirm that the bulk bands remain gapless.
However, a key characteristic of a semimetal is the crossover of the bulk bands. We plot the bulk band structure of the topological insulator film with $d$-wave altermagnet in Fig.~\ref{fig3}(c).
It is clear that while the Weyl points at $X_1$ and $X_2$ become gapped, the Weyl points at $\Gamma$ and $M$ persist.
The positions of these remaining Weyl points in the first Brillouin zone are indicated by cyan markers in Fig.~\ref{fig3}(d).
The simultaneous presence of corner states in finite-size nanoflakes and the Weyl points in the bulk band directly signals the realization of 2D higher-order Weyl semimetals.

To understand the physical origin of the corner states and the persistence of Weyl points, we analyze the symmetry properties of the system.
Although the $d$-wave altermagnet breaks time-reversal symmetry and the $\mathcal{M}$ symmetry, leading to gaps in both the helical edge states and the Weyl points at $X_1$ and $X_2$, the Hamiltonian along the high-symmetry directions $k_x = k_y$ and $k_x = -k_y$ retains $\mathcal{M}$ symmetry.
Specifically, $H(k_x, k_x)$ and $H(k_x, -k_x)$ remain invariant under the $\mathcal{M}$ operator.
In the three subspaces defined by $\mathcal{M}$, the Hamiltonian $H(k_x, k_x)$ can be decomposed into three independent components:
\begin{align}
H_{\pm 1}&= v_F \sin k_x (\tau_x \mp \tau_y) + \sqrt{2} m(k) \tau_z
\equiv \textbf{d}_{\pm1}(k)\cdot {\bf{\tau}},\nonumber \\
H_{0}&= v_F \sin k_x (\tau_x - \tau_y).
 \label{eq5}
\end{align}
Clearly, the subspace Hamiltonians $H_{-1}$, $H_{1}$, and $H_{0}$ all exhibit chiral symmetry. The Hamiltonians $H_{0}$ has two gapless points in momentum space at $k_x = 0$ and $k_x = \pi$, corresponding to the Weyl points located at the $\Gamma$ and $M$ points.
For the bulk gapped Hamiltonians $H_{-1}$ and $H_{1}$, we calculate their winding numbers using the following expression:
\begin{align}
  v_{\pm 1}=\frac{1}{2 \pi} \int_{-\pi} ^\pi (\textbf{d}_{\pm1}(k) \times \frac{d}{dk} \textbf{d}_{\pm1}(k) )d k_x.
 \label{eq6}
\end{align}
The calculated winding numbers are $v_{\pm 1} = \pm 1$, which indicates the presence of two corner states in the 2D finite-size system, located at the endpoints of the $k_x = k_y$ path.
For the case of $k_x = -k_y$, the Hamiltonian $H(k_x, -k_x)$ can similarly be decomposed into three distinct components:
\begin{align}
H_{\pm 1}&= v_F \sin k_x (\tau_x \pm \tau_y) + \sqrt{2} m(k) \tau_z,\nonumber \\
H_{0}&= v_F \sin k_x (\tau_x + \tau_y).
 \label{eq7}
\end{align}
The winding numbers $v_{\pm 1} = \mp 1$ are calculated for $H_{-1}$ and $H_{1}$, confirming the presence of two corner states at the endpoints of the $k_x = -k_y$ path.
This analysis demonstrates that the system is a 2D higher-order Weyl semimetal, characterized by bulk Weyl points and corner states.
These findings are in excellent agreement with the numerical results presented in Fig.~\ref{fig3}.

\begin{figure}
  \centering
  \includegraphics[width=8.5cm,angle=0]{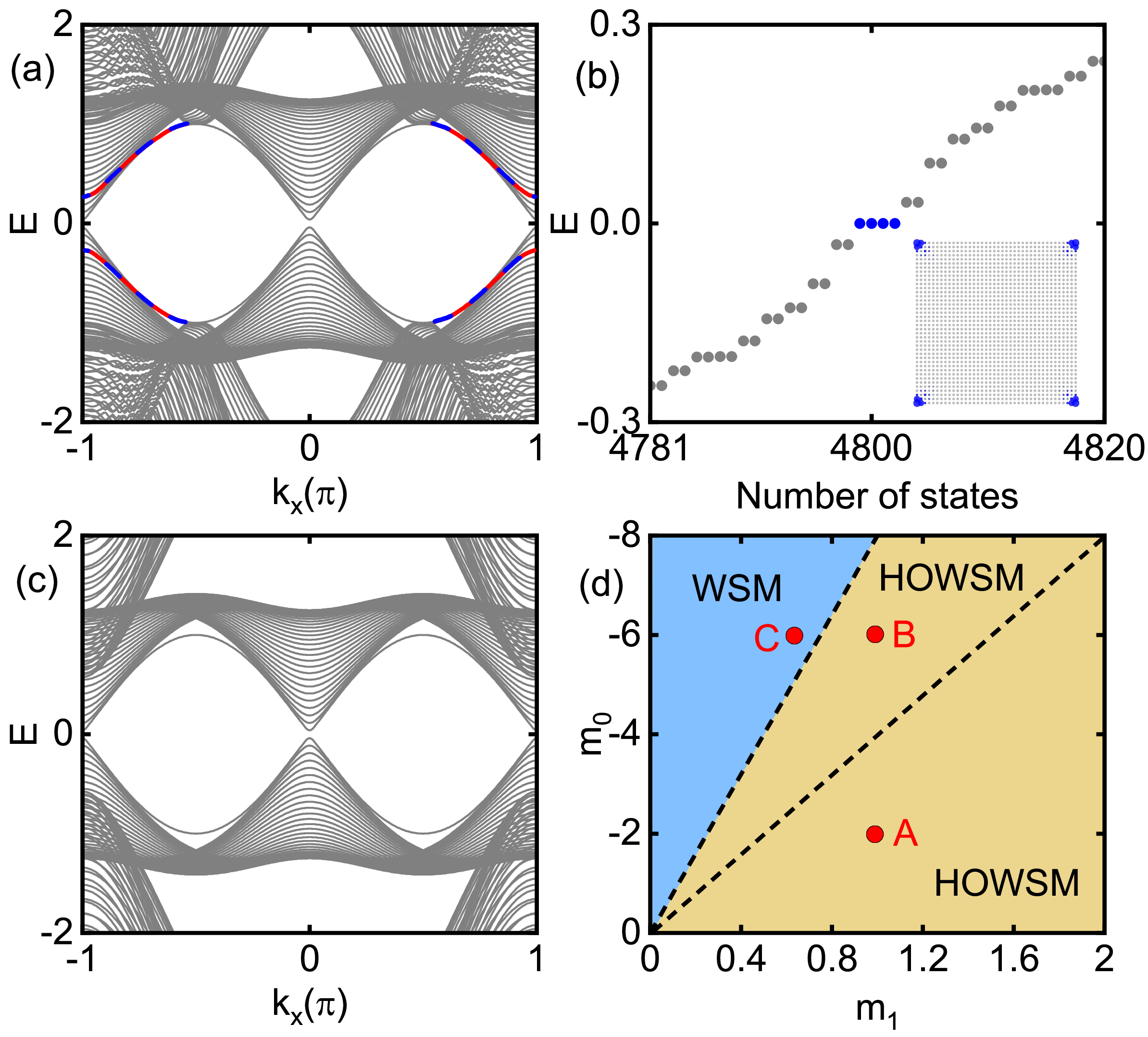}
\caption{(a) The energy bands correspond to the parameters $v_F = 1, m_0=-6, m_1=1$, and $J=0.6$.
   The blue solid lines and the red dashed lines indicate the gapped helical edge states.
   (b) Energy levels of the rectangular-shaped nanoflake. Blue dots correspond to corner states. The wave function probability distribution of the corner states is shown in the inset (blue density dots).
   (c) The energy bands correspond to the parameters $v_F = 1, m_0=-6, m_1=0.7$, and $J=0.6$.
  (d) Topological phase diagram of the trilayer film as functions of $m_1$ and $m_0$ with $v_F = 1$, and $J=0.6$.
    The black dashed lines indicates the phase boundaries ($|m_0| = 8 |m_1|$ and $|m_0| = 4 |m_1|$). 
    HOWSM corresponds to a higher-order Weyl semimetal phase with corner states, WSM for normal Weyl semimetal. Parameter point coordinates $(m_1,m_0)$ : $A (1,-2)$, $B(1,-6)$, and $C(0.7,-6)$.
}
\label{fig4}
\end{figure}

\subsection{Topological phase diagram}

In the above analysis, we have chosen specific parameters to analyze the topological phases.
In this section, we systematically explore the effect of varying parameters on the topological phase of the system.
We first adjust the value of $m_0$, setting the system parameters to $v_F=1, m_0=-6, m_1=1,$ and $J=0.6$.
Interestingly, as shown in the energy band structure in Fig.~\ref{fig4}(a), the gapped helical edge states now appear at $k_x = \pi$, which is different from the case shown in Fig.~\ref{fig3}(a).
To understand the topological properties of the gapped edge states, we plot the energy levels of the rectangular-shaped nanoflake in Fig.~\ref{fig4}(b).
The inset displays the wavefunction distributions of the four zero-energy states (highlighted in blue), uniformly distributed over the four corners, indicating a two-dimensional higher-order topological phase.
Although the system remains in the higher-order Weyl semimetal phase both before and after adjusting $m_0$, the edge states shift from $k_{x}=0$ to $k_{x}=\pi$), suggesting that a topological phase transition has occurred.
We next adjust $m_1$ to analyze the system in more detail, with the system parameters set to $v_F=1, m_0=-6, m_1=0.7,$ and $J=0.6$. 
The energy band structure of the nanoribbon is shown in Fig.~\ref{fig4}(c), where the helical edge state disappears, and the system behaves as a normal Weyl semimetal phase.

To clearly understand and determine the phase boundaries, we analyze the Hamiltonian of the symmetric subspace in Eq.~(\ref{eq7}).
It is clear that no matter how $m_1$ and $m_0$ vary, a bulk band of $H_0$ always closes, giving rise to two Weyl points in the trilayer system.
The bulk band closure (or phase boundary) of $H_{\pm 1}$ corresponds to $m_0 m_1 = 0$ or $ |m_0| = 8 |m_1|$ or $|m_0| = 4 |m_1|$, dividing the phase diagram in Fig.~\ref{fig4}(d) into three distinct regions.
Using Eq.~(\ref{eq6}), we calculate the winding number of $H_{\pm 1}$ and find that the winding number $v_{\pm 1} = \mp 1$ while $m_0 m_1 < 0$ and $ |m_0| < 8 |m_1|$ (excluding $|m_0| =4 |m_1|$), while $v_{\pm 1} = 0$ in all other cases.
Therefore, the yellow region corresponds to the higher-order Weyl semimetal phase, and the blue region corresponds to the normal Weyl semimetal phase. This is consistent with the numerical results.
The parameter points $A$, $B$, and $C$ correspond to the energy band structures in Figs.~\ref{fig3}(a),~\ref{fig4}(a), and~\ref{fig4}(c), respectively.
Since the $d$-wave altermagnet still preserves the Weyl points in the bulk and only opens a gap in the helical edge state, the small altermagnet does not affect the topological phase diagram.

\section{Conclusions}

We have demonstrated that an out-of-plane $d$-wave altermagnet can transform a trilayer topological insulator film into a 2D higher-order topological Weyl semimetal, featuring corner states.
Specifically, the pristine trilayer topological insulator film functions as a 2D Weyl semimetal, exhibiting helical edge states and four bulk Weyl points.
The systems's topological properties are characterized by the Chern number and the winding numbers of symmetric subspaces.
Introducing an out-of-plane $d$-wave altermagnet on the top and bottom layers induces a gap in the helical edge states while preserving two bulk Weyl points.
We have further analyzed the energy levels of a rectangular nanoflake, which confirms the existence of corner states.
The emergence of these corner states is attributed to the nonzero winding number along the symmetric subspace of the high-symmetry line $k_x = \pm k_y$.
Finally, we provide a complete second-order topological phase diagram.
Our findings provide compelling insights into the realization of pioneering 2D higher-order Weyl semimetals.

We also point out that placing a Bi$_2$Se$_3$ trilayer topological insulator film sandwiched between two altermagnetic material MnF$\rm _2$~\cite{Yuan2020, Smejkal2020}, with the magnetic proximity effect~\cite{Vobornik2011} inducing altermagnetism in the top and bottom layers, promises to achieve a 2D higher-order Weyl semimetal.
Experimentally, charged topological corner states can be detected using scanning-tunneling microscopy (STM)~\cite{Yin2021}.
When the STM probe is at the energy of the other states, a broadening peak can be observed.
After the corner states are induced by altermagnet, sharp peaks can be observed at the corners of the sample with STM.
Therefore, observing changes in the energy spectrum peaks with STM can be a strong evidence for the existence of corner states.

\section*{Acknowledgements}

This work was financially supported by the National Natural Science Foundation of China (Grants No. 12074097, No. 12374034, and No. 11921005),
Natural Science Foundation of Hebei Province (Grant No. A2024205025),
the National Key R and D Program of China (Grant No. 2024YFA1409002),
and the Innovation Program for Quantum Science and Technology (Grant No. 2021ZD0302403).

\end{document}